\documentclass[journal]{IEEEtran}
\usepackage[utf8]{inputenc}
\usepackage{amsmath,amssymb,amsthm}
\usepackage{gensymb}

\usepackage{subcaption}
\usepackage{xspace,arydshln}
\usepackage[normalem]{ulem}

\usepackage{graphicx,fancyhdr,epstopdf}
\usepackage[ruled]{algorithm2e}
\usepackage{enumitem}
\usepackage{authblk}
\usepackage{color}
\usepackage{cite}
\usepackage{hyperref}
\usepackage{soul}
\usepackage{etoolbox}

\definecolor{lee}{rgb}{0,0.8,0}
\definecolor{dblue}{rgb}{0,0,0.8}

\newcommand{\stkout}[1]{\ifmmode\text{\sout{\ensuremath{#1}}}\else\sout{#1}\fi}

\begin{document}

\title{Massive MIMO Channel Estimation with Low-Resolution Spatial Sigma-Delta ADCs}
\author{Shilpa~Rao,~\IEEEmembership{Student~Member,~IEEE,}
        Gonzalo~Seco-Granados,~\IEEEmembership{Senior~Member,~IEEE,}
        Hessam~Pirzadeh,~\IEEEmembership{Student~Member,~IEEE,}
        Josef~A.~Nossek,~\IEEEmembership{Life~Fellow,~IEEE,}
        and~A.~Lee~Swindlehurst,~\IEEEmembership{Fellow,~IEEE}
\thanks{This work was supported by the U.S. National Science Foundation under Grants CCF-1703635 and ECCS-1824565.}
}
\maketitle
\begin{abstract}
We consider channel estimation for an uplink massive multiple-input multiple-output (MIMO) system where the base station (BS) uses an array with low-resolution (1-2 bit)
analog-to-digital converters and a spatial Sigma-Delta ($\Sigma\Delta$) architecture
to shape the quantization noise away from users in some angular sector. We develop a linear minimum mean squared error (LMMSE) channel estimator based on the Bussgang decomposition that 
reformulates the nonlinear quantizer model using an equivalent linear model plus quantization noise. We also analyze the uplink achievable rate with maximal ratio combining (MRC), zero-forcing (ZF) and LMMSE receivers and provide a lower bound for the achievable rate with the MRC receiver. Numerical results show superior channel estimation and sum spectral efficiency performance using the $\Sigma \Delta$ architecture compared to conventional 1- or 2-bit quantized massive MIMO systems.
\end{abstract}

\section{Introduction}
Massive MIMO systems provide high spatial resolution and throughput, but the 
cost and power consumption of the associated RF hardware, particularly the analog-to-digital converters (ADCs) and digital-to-analog converters (DACs) can be 
prohibitive, especially at higher bandwidths and sampling rates. To save power and chip area, low-resolution quantizers have been suggested. The simple design and low power consumption of low-resolution ADCs/DACs have made them very suitable for massive MIMO. 

There has been extensive research on one-bit data converters, in particular, for uplink transmission~\cite{Li_channel,Jacobs,
mollen217tpc,Choi_TPC, jeon2018one,dong2017performance} and downlink precoding~\cite{mezghani2009transmit,li2017downlink,saxena2017analysis,jedda2018quantized,landau2017branch}. Channel estimation based on generalized approximate message passing~\cite{garcia2016channel}, near maximum likelihood~\cite{Choi_TPC} and linear approaches~\cite{Li_channel} using one-bit ADCs, support vector machines~\cite{LyS21_SVM}, recursive least-squares using two-bit ADCs~\cite{mehanna2014channel} and with dithered feedback signals~\cite{dabeer2010channel} using 1-3 bit ADCs have been proposed. Some recent works~\cite{Choi_globecom,Hong_MDD,LyS20,LyS21_DNN} study non-linear and learning-based detection methods foregoing the channel estimation stage. In addition to savings in energy and circuit complexity, one-bit ADCs also reduce the fronthaul throughput from the remote radio head (RRH) in cellular applications. While it has been shown that one-bit quantization causes only a small degradation at low SNRs and offers significant advantages in terms of power consumption and implementation complexity, at medium to high SNRs the performance loss is substantial.

One method to improve the performance is to simply increase the number of quantization bits. Simulations have shown that using ADCs with 3-5 bits of resolution in massive MIMO provides 
performance that is very close to that achievable with infinite precision, and achieves a good trade-off between energy and spectral efficiency~\cite{Roth_hybrid,pirzadeh2017spectral}. Alternatively, the sampling rate at which the one-bit quantizers operate can also be increased~\cite{koch_capacity,shamai1994information,gok2017}.
In~\cite{Ucuncu_2018,Shao_1bitover,stein2017performance,schluter2018,landau20171}, parameter estimation with oversampled one-bit receivers has been analyzed, and downlink precoding has been studied in~\cite{melo2020zero}. A well-known technique that combines one-bit quantization and oversampling is the  $\Sigma\Delta$ ADC, which to date has primarily found application in 
ultrasound imaging, automotive radar and pulse-coded modulation for audio encoding. The temporal $\Sigma\Delta$ converter scheme consists of an oversampled modulator, responsible for
digitization of the analog signal, followed by a negative feedback loop that shapes the quantization noise with a simple high-pass filter. The quantization noise can then be removed in favor of the desired signal using a digital lowpass filter and decimation. The temporal $\Sigma\Delta$ architecture has been extensively studied~\cite{aziz1996overview,gray1987oversampled,Gray_sine}, and higher-order implementations exist that can provide additional frequency-selective noise shaping. The use of $\Sigma\Delta$ ADCs in parallel architectures for MIMO systems has been studied in~\cite{palguna2016,Venk2011}. The Bussgang theorem~\cite{bussgang1952crosscorrelation} is used to analyze the noise shaping effect of temporal $\Sigma\Delta$ modulators, and Price's theorem~\cite{price1958useful} is used in~\cite{kaveh_temporal} to predict the noise floor. While higher-resolution ADCs and temporal oversampling can improve performance with only a moderate increase in power consumption, they significantly increase the required fronthaul throughput compared with one-bit quantization.

A noise-shaping effect analogous to that achieved by temporal $\Sigma\Delta$ quantizers can be achieved by implementing the $\Sigma\Delta$ architecture in the spatial domain. This approach exploits oversampling in space, which can occur in massive MIMO settings with a limited array aperture, or for scenarios where the uplink signals are confined to a given angular sector, due to cell sectorization 
or certain small-cell geometries (narrow conference halls, city streets, etc.). In a system employing spatial $\Sigma\Delta$ ADCs, the quantization error from the ADC at one antenna is fed to the input of an adjacent antenna, rather than to the input of the same antenna. In a standard implementation without phase-shifting the feedback, this has the effect of shaping the quantization noise to high spatial frequencies, in favor of signals that arrive from low spatial frequencies (e.g., angles closer to the broadside of a uniform linear array). While spatial oversampling, i.e., antennas spaced less than one-half wavelength apart, can produce the required low spatial frequencies for the user signals, there is a limit to how close the antennas can be placed together before mutual coupling and the physical size of the antennas come into play. The effect of mutual coupling becomes more prominent as the inter-element spacing is decreased below one-half wavelength, and can degrade the estimation performance~\cite{Lin_blind}. For this reason, a likely scenario for the $\Sigma\Delta$ array concept is that some degree of spatial oversampling would be combined with the assumption that the users of interest are already confined to some angular sector, since one can easily steer the angular sector of low quantization noise to arbitrary directions.

While temporal $\Sigma\Delta$ systems have been studied for decades, there is relatively little work on corresponding spatial implementations. Only recently has the noise shaping characteristics of first and second-order spatial and cascaded space-time $\Sigma\Delta$ architectures been studied for a few array processing applications. In particular, applications have been considered for massive MIMO~\cite{Corey_Sig,baracspatial,rao2019massive,Shao_precoding,pirzadeh2019spectral}, phased arrays~\cite{scholnik2004spatio,kriegerDense}, interference cancellation~\cite{Venk2011}, and spatio-temporal $\Sigma\Delta$ circuit implementations~\cite{madan2017,nikoofard}. With the exception of our preliminary studies in~\cite{rao2019massive,channel20asilomar}, there has been no prior work focused on channel estimation for spatial $\Sigma\Delta$ massive MIMO systems. 

In this paper, we consider optimal linear minimum mean squared error (LMMSE) channel estimation for massive MIMO systems with first-order one-bit and two-bit spatial $\Sigma\Delta$ ADCs. Our initial results on this problem in~\cite{rao2019massive} were derived based on a vector-wise Bussgang decomposition similar to what was used for standard one-bit quantization in~\cite{Li_channel}. However, this approach leads to a mathematical model in which the quantization error vector is defined to be uncorrelated with the input vector to the $\Sigma\Delta$ quantizer, which is not consistent with the more traditional definition of quantization noise. Based on this observation, we have modified our analysis to incorporate a more meaningful definition of the quantization noise, using an element-wise implementation of the Bussgang decomposition as defined in~\cite{pirzadeh2019spectral} in order to find an equivalent linear signal-plus-quantization-noise representation. Our approach explicitly takes into account the spatial correlation between the quantized outputs of the $\Sigma\Delta$ ADC array. Our channel model also includes the impact of mutual coupling between the antenna elements, including also the fact that the noise becomes spatially correlated when the receiver antennas are closely spaced.

{\color{black} Use of the Bussgang decomposition for channel estimation with $\Sigma\Delta$ ADCs is considerably more difficult than the case for standard quantization considered in \cite{Li_channel}, due to the presence of error feedback to adjacent antennas. One of our key contributions is to show how the structure of the $\Sigma\Delta$ array can be exploited} to find a recursive solution for the covariance matrices required to compute the LMMSE channel estimate, {\color{black} and we derive a practical algorithm for doing so. We also derive analytical expressions for the resulting covariance matrix of the channel estimation error}. After performing the analysis for the one-bit case, we show how the analysis can be extended for a $\Sigma\Delta$ array implemented with two-bit quantization, {\color{black} and similar extensions are possible for higher resolution ADCs}. The resulting estimators have low complexity and the simulated estimation error closely matches the derived analytical expressions. Our simulation results indicate that, at low-to-medium SNRs, the LMMSE channel  estimator for the $\Sigma\Delta$ array yields channel estimates that are very close to those achieved with infinite resolution ADCs. The inevitable error floor at high SNR is significantly lower than that achieved by standard one-bit and two-bit quantization; for example, the one-bit $\Sigma\Delta$ LMMSE channel estimate has an {\color{black}error} that is two orders of magnitude smaller than that achieved by standard two-bit quantization.

The spectral efficiency of one-bit $\Sigma\Delta$ arrays was recently analyzed in~\cite{pirzadeh2019spectral,pirzadeh_spawc} {\color{black} with and without mutual coupling, respectively}, but only for the case where the channel is perfectly known. {\color{black} As our second primary contribution}, we extend the analysis of~\cite{pirzadeh2019spectral,pirzadeh_spawc} to derive a lower bound on the uplink achievable rate using the maximal ratio combining (MRC), zero-forcing (ZF) and LMMSE receivers when implemented with imperfect channel state information (CSI) obtained using our LMMSE channel estimate, {\color{black} assuming the detailed model for mutual coupling in \cite{pirzadeh_spawc}}. We derive a closed-form expression for the spectral efficiency of MRC, while for the ZF and LMMSE receivers we present simulated results. In both cases, the results of our spectral efficiency analysis show that the $\Sigma\Delta$ architecture significantly outperforms standard low-resolution (1-2 bits) quantization. For the ZF and LMMSE receivers, the $\Sigma\Delta$ implementation achieves twice the spectral efficiency of standard one- and two-bit approaches. For the case of MRC, the two-bit $\Sigma\Delta$ architecture provides 99\% of the spectral efficiency of a system with infinite resolution ADCs. The results presented in the following sections go beyond the preliminary work in \cite{rao2019massive,channel20asilomar} by providing details of the derivations of the channel estimator and sum spectral efficiency expressions, incorporation of mutual coupling and correlated receiver noise in the channel model, and including analysis of the LMMSE receiver performance. 

{\em Notation}: Boldface lowercase variables denote vectors and boldface uppercase variables denote matrices. ${\bf X}^T$, ${\bf X}^H$ and ${\bf X}^*$ are the transpose, Hermitian transpose, and conjugate of ${\bf X}$, respectively. The $i$th element of vector ${\bf x}$ is represented by ${x}_i$. The symbols $\otimes$ and ${\rm vec}(\cdot)$ respectively represent the Kronecker product and the vectorization operation, i.e., the stacking of the columns of a matrix one below the other. Real and imaginary parts are indicated by ${\rm Re}(\cdot)$ and ${\rm Im}(\cdot)$, respectively. $\mathbb{E}[\cdot]$ is the expectation operator. ${\rm Tr}\left( {\bf X} \right)$ is the trace of the matrix ${\bf X}$ and ${\bf X}^{\dagger}$ denotes pseudo-inverse of ${\bf X}$. The matrix ${\bf I}_p$ denotes a $p \times p$ identity matrix. The function $\mbox{\rm mod}_M(\cdot)$ represents the modulo-$M$ operator, and $\lfloor z \rfloor$ is the largest integer smaller than $z$. A circularly symmetric complex Gaussian vector with mean ${\bf a}$ and covariance matrix ${\bf B}$ is denoted by ${\bf x} \sim \mathcal{CN}\left( {\bf a}, {\bf B} \right)$. The cumulative distribution function (cdf) and the standard normal density are given by $\Psi(x)$ and $\Psi^{\prime}(x)$, respectively. ${\rm Ci}(x) = \eta + {\rm log}(x) + \int_0^x \frac{{\rm cos} t -1}{t} dt$ and ${\rm Si}(x) = \int_0^x \frac{{\rm sin} t}{t} dt$ are the cosine and sine integral functions, respectively, where $\eta$ is the Euler–Mascheroni constant.

\section{System Model}
\label{SysModel}
We consider an uplink massive MIMO system with $K$ single-antenna user 
terminals, and a base station (BS) equipped with a uniform linear array (ULA) of $M$ antennas and a first-order spatial $\Sigma\Delta$ array. During the training period, all $K$ users simultaneously transmit their pilot sequences of length $N$. The received signal, ${\bf X} \in \mathbb{C}^{M \times N}$, at the BS is
\begin{equation}
 {\bf X} = \left[ {\bf{x}}_1 \; \cdots \; {\bf{x}}_N \right] = \sqrt{\rho} \,{\bf G}\, \boldsymbol{\Phi}_t + {\bf N},
 \label{Eq1}
\end{equation}
where ${\bf{x}}_k$ is the array output for training sample $k$, ${\bf G} \in 
\mathbb{C}^{M \times K}$ is the channel matrix, and $\boldsymbol{\Phi}_t \in 
\mathbb{C}^{K \times N}$ is the pilot matrix. The matrix ${\bf N} = \left[{\bf n}_1, {\bf n}_2, \dots, {\bf n}_N \right]$ contains noise from both intrinsic (low-noise amplifiers and other circuitry) and extrinsic (received by the antennas) sources and consists of zero mean spatially correlated additive noise Gaussian elements that satisfy
\begin{equation}
\begin{aligned}
\mathbb{E}\left[ {\bf n}_i {\bf n}_i^H \right] &= {\bf C}_N,\\
\mathbb{E}\left[ {\bf n}_i {\bf n}_j^H \right] &= {\bf 0}, \quad i \neq j.
\end{aligned}
    \label{Eq1a}
\end{equation}
{\color{black} For simplicity, we assume that power control is employed such that} all the user signals are received with the same power and, therefore, $\rho$ is a factor that determines the SNR, 

We consider a spatially correlated channel model for ${\bf G}$. {\color{black} In particular, the $k$th column of ${\bf G}$, which represents the channel ${\bf g}_k$ for the $k$th user, is assumed to be given by
\begin{equation}
{\bf g}_k = {\bf C}_{g_k}^{\frac{1}{2}} {\bf h}_k \; ,
\label{Eq2}
\end{equation}
where the elements of ${\bf h}_k \in \mathbb{C}^{L_k \times 1}$ are independently and identically distributed (i.i.d.) as $\mathcal{CN}(0,1)$ random variables, and ${\bf C}_{g_k} \in \mathbb{C}^{M \times M}$ is the channel covariance matrix. In our application, the spatial correlation will arise from two factors: signal arrivals that come from a certain sector of possible angles of arrival (AoAs), and mutual coupling between the BS antennas:
\begin{equation}
{\bf C}_{g_k}^{\frac{1}{2}} = \frac{1}{\sqrt{L_k}} {\bf T} {\bf A}_k \; , 
\label{CGk}
\end{equation}
where ${\bf T}$ is a matrix that accounts for the mutual coupling, ${\bf A}_k \in \mathbb{C}^{M \times L_k}$ is a matrix whose $l$th column is the steering vector of a linear array
\begin{equation}
    {\bf a}_l = \left[1 \; \; e^{-j2 \pi d_{12} {\rm sin}(\theta_l)} \; \cdots \; e^{-j2 \pi d_{1M} {\rm sin}(\theta_l)} \right]^T \; ,
\end{equation}
$\theta_l$ is the AoA, and $d_{ij}$ is the distance between the antenna elements $i$ and $j$. We will assume that the AoAs for all users lie in a certain known angular region $\theta_l \in \mathcal{S}_{\theta}$. This is common situation in many practical settings due to cell sectoring; {\em e.g.,} a given BS array will only serve users from some fraction of all available azimuth or elevation angles. Besides this region $\mathcal{S}_{\theta}$, prior to channel estimation the BS is only aware of the channel covariance matrix ${\bf C}_{g_k}$ for each user, and not the components (mutual coupling, AoAs) of its decomposition in\eqref{CGk}. 

The impact of mutual coupling is important because we will be considering arrays whose elements may be spaced closer than one-half wavelength apart.} In modeling the mutual coupling, the antenna array is treated as a bilateral network and the relationship between element output voltages and open circuit voltages is derived from multiport circuit theory. The analytical formulas involved in the expression for ${\bf T}$ are detailed in~\cite{schelkunoff1952,ivrlavc2010toward,pirzadeh_spawc} and summarized below. The matrix ${\bf T}$ is given by
\begin{equation}
    {\bf T} = \left( {\bf I} + \frac{1}{R} {\bf Z}\right)^{-1},
\end{equation}
where $R$ is the input impedance of the low-noise amplifier (LNA). The impedance matrix of the antenna elements, ${\bf Z}$, is described by 
\begin{equation}
\begin{aligned}
    &{\bf Z}_{ij} = 30 \left( 2{\rm Ci}(2\pi d_{ij}) - {\rm Ci}(\xi_{ij} + \pi) - {\rm Ci}(\xi_{ij} - \pi) \right. \\
  & \left. \hspace{1cm} - j \left(2{\rm Si}(2\pi d_{ij}) - {\rm Si}(\xi_{ij} + \pi) - {\rm Si}(\xi_{ij} - \pi) \right) \right), i \neq j\\
    &{\bf Z}_{ii} = 30\left( \eta + {\rm log}(2\pi)- {\rm Ci}(2\pi) + j{\rm Si}(2\pi)  \right),
\end{aligned}
\end{equation}
where $\xi_{ij} = \pi \sqrt{1+4 d_{ij}^2}$. 

Furthermore, due to the presence of mutual coupling, the noise at the receiver is spatially correlated and its covariance matrix is given by 
\begin{eqnarray}
   {\bf C}_N & = & {\bf T} \boldsymbol{\Upsilon} {\bf T}^H, \\
   \boldsymbol{\Upsilon} & = & \sigma_i^2  \left( {\bf Z}{\bf Z}^H -2 R_N {\rm Re}(\varrho_n^*{\bf Z}) + R_N^2 {\bf I} \right) \\ & & \qquad + 4k_B T_A B {\rm Re}({\bf Z}) \; ,
\end{eqnarray}
where $\sigma_i^2 = \mathbb{E}[{\bf i}_N {\bf i}_N^H]$, $\sigma_v^2 = \mathbb{E}[{\bf v}_N {\bf v}_N^H]$, ${\bf i}_N$ and ${\bf v}_N$ are the complex current and voltage of the noise source, $R_N = \sigma_v/\sigma_i$ is the noise resistance, {\color{black}$\varrho_n= \mathbb{E}[{\bf i}_N {\bf v}_N^H]/(\sigma_i \sigma_v) $} is the so-called noise correlation coefficient, $k_B$ is the Boltzmann constant, $T_A$ is the ambient temperature, and $B$ is the bandwidth. 

Vectorizing~(\ref{Eq1}), we get
\begin{equation}
\begin{aligned}
 {\bf x} = {\rm vec} \left( {\bf X} \right) = \boldsymbol{\Phi} {\bf g} + {\bf n},
 \end{aligned}
 \label{Eq3}
\end{equation}
where
\begin{equation}
\begin{aligned}
 \boldsymbol{\Phi} &= \sqrt{\rho} \left( \boldsymbol{\Phi}_t^T \otimes {\bf I}_{M} \right) \\
{\bf g} & = \left[ \begin{array}{c}
     {\bf g}_1  \\
      {\bf g}_2 \\
      \vdots \\
      {\bf g}_K
\end{array} \right]= {\rm 
vec}\left( {\bf G} \right) \\
{\bf n} &= \left[ \begin{array}{c}
     {\bf n}_1  \\
      {\bf n}_2 \\
      \vdots \\
      {\bf n}_N
\end{array} \right] = {\rm vec} \left( {\bf N} \right).
\end{aligned}
\end{equation}
The covariance matrix of ${\bf x}$ can be expressed as
\begin{equation}
{\bf C}_x = \boldsymbol{\Phi} {\bf C}_g \boldsymbol{\Phi}^H + {\bf C}_{n}, 
\end{equation}
{\color{black}where we have defined the block-diagonal matrices ${\bf C}_g = {\rm blkdiag}\{{\bf C}_{g_1},\dots,{\bf C}_{g_K} \}$ and ${\bf C}_{n} = \mathbf{I}_{N} \otimes {\bf C}_N$. The average per-user per-antenna SNR is defined as}
\begin{equation}
    {\rm SNR} = \frac{\rho}{NK} \frac{{\rm Tr}\left( \mathbb{E} \left[ {\bf G} \boldsymbol{\Phi}_t \boldsymbol{\Phi}_t^H {\bf G}^H \right] \right)}{{\rm Tr}\left( {\bf C}_N \right)}.
\end{equation}
When the pilot sequences are row-wise orthogonal and the minimum number of pilots, $N=K$, is used, then $\boldsymbol{\Phi}_t \boldsymbol{\Phi}_t^H = K {\bf I}_K$ and 
{\color{black}\begin{equation}
    {\rm SNR} = \frac{\rho}{K} \frac{{\rm Tr}\left( {\bf C}_g \right)}{{\rm Tr}\left( {\bf C}_N \right)}.
\end{equation}}

In the derivations below, we will assume that the power of the pilot signals is time-invariant, which implies that the diagonal elements of $\boldsymbol{\Phi}_t^H \boldsymbol{\Phi}_t$ are identical. The goal of the paper is to derive an algorithm to estimate the channel ${\bf g}$ from the output ${\bf x}$ of the $\Sigma\Delta$ array, and this algorithm is presented below in Section~\ref{ChEst}. Prior to development of the algorithm, we first present details about the $\Sigma\Delta$ array architecture in Section~\ref{SpatialSD}, and derive an equivalent linear model for the architecture in Section~\ref{LinMod} for the case of both one-bit and two-bit ADCs.

\color{black}
\section{First-Order Spatial $\Sigma \Delta$ Architecture}
\label{SpatialSD}

{\color{black} The discussion in the next two sections parallels that in \cite{pirzadeh2019spectral}, but it includes some important distinctions that are necessary for the channel estimation problem, as well as an extension to the case of two-bit ADCs.}
The first-order temporal $\Sigma\Delta$ modulator consists of a low-resolution quantizer $\mathcal{Q}(\cdot)$ and a negative feedback loop where the difference between the output and input of the quantizer is subtracted from the antenna input $x[n]$ after a one-sample delay. The operation of the quantizer is defined using the following equivalent linear model with gain $\gamma > 0$:
\begin{equation}
\label{equivlin}
    y[n] = \mathcal{Q}\left(r[n]\right) = \gamma r[n]+q[n] \; .
\end{equation}
While the gain of the quantizer is normally taken to be simply $\gamma=1$, we consider this more general case since it is relevant to our subsequent analysis.  As shown in~\cite{pirzadeh2019spectral}, this leads to the following transfer function description of the $\Sigma\Delta$ modulator:
\begin{eqnarray*}
    Y(z) & = & \frac{\gamma}{1-(\gamma-1)z^{-1}} X(z) + \frac{(1-z^{-1})}{1-(\gamma-1)z^{-1}} Q(z) \\[6pt]
    & = & A_x(z) X(z) + A_q(z) Q(z) \; ,
\end{eqnarray*}
where $\{Y(z), X(z), Q(z)\}$ respectively represent the $z$-transforms of $\{y[n], x[n], q[n]\}$. When $\gamma=1$, we have that $A_x(z)=1$ is an all-pass filter and that $A_q(z)=1-z^{-1}$ is a first-order high-pass filter, which is the standard result, indicating the quantization noise is shaped to higher frequencies. Given that $x[n]$ is oversampled and is concentrated at lower frequencies, the effect of the quantization noise can be substantially reduced by passing the output of the $\Sigma\Delta$ modulator through a low-pass filter. The all-pass plus high-pass structure still remains true as long as $\gamma \approx 1$, but the $\Sigma\Delta$ modulator approaches instability as $\gamma \rightarrow 2$.

\begin{figure}
  \includegraphics[width=3in]{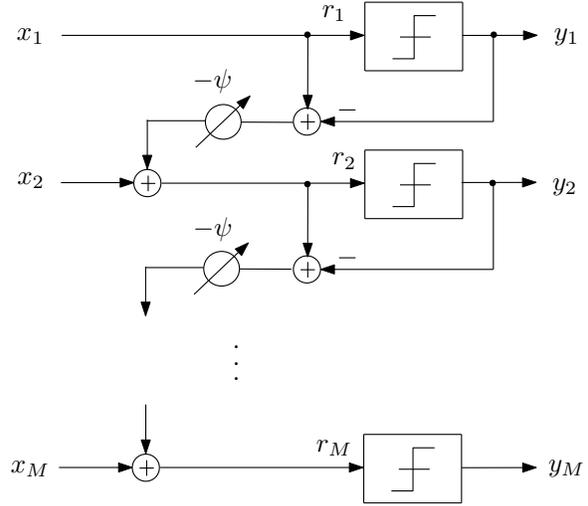} 
\caption{A first-order one-bit $\Sigma\Delta$ array steered to direction $\psi$.}
\label{fig:fig1}
\end{figure}

Quantization noise shaping can also be achieved in the spatial domain by propagating the quantization error from one antenna to the next, in effect exploiting spatial rather than temporal correlation in the desired signal. The spatial $\Sigma\Delta$ architecture is depicted in Fig.~\ref{fig:fig1}, and shows that the quantization error from one antenna is phase-shifted by $-\psi$ prior to being added to the input of the adjacent antenna. This architecture shapes the quantization noise away from the angle of arrival (AoA) associated with the phase shift $\psi$, and thus users in an angular sector surrounding this AoA experience a significantly higher signal-to-quantization-noise ratio (SQNR). The size of the high-SQNR angular sector can be increased by placing the antennas closer together than $\lambda/2$, corresponding to spatial oversampling, although in practice mutual coupling and the physical dimensions of the antennas place a limit on how much spatial oversampling can be achieved.

To generate a mathematical model for the $\Sigma\Delta$ array, define the $MN\times 1$ vectors $\bf{r}$ and $\bf{y}$ corresponding respectively to the quantizer inputs and the array outputs that result from the $MN\times 1$ received signal vector $\bf{x}$ in~\eqref{Eq3}. In other words, the $m$th elements $\{x_m, r_m, y_m\}$ of the vectors $\{\bf{x},\bf{r},\bf{y}\}$ respectively represent the signal received by antenna $\mbox{\rm mod}_M(m)$, the input to the quantizer at antenna $\mbox{\rm mod}_M(m)$, and the $\Sigma\Delta$ output of antenna $\mbox{\rm mod}_M(m)$ for training sample $\lfloor \frac{m}{M} \rfloor$. Defining $m'=\mbox{\rm mod}_M(m)$, we thus have
\begin{equation}
 y_m = \alpha_{m'}  \mathcal{Q}_{m'} \left( {\rm Re}(r_m) \right) +  j \alpha_{m'} \mathcal{Q}_{m'} \left( {\rm Im}(r_m) \right),
 \label{Eq4}
\end{equation}
where $\mathcal{Q}_{m'}$ represents the quantization operation for antenna $m'$, {\color{black} and $\alpha_{m'}$ is the output level of the quantizer. We allow $\alpha_{m'}$ to be different for each antenna, unlike conventional one-bit quantization where they are the same}. In vector form, the output of the $\Sigma\Delta$ array can be written as
\begin{equation}
 \begin{aligned}
  {\bf y} &= \mathcal{Q} ( {\bf r} ) \\
  &= \left[ \mathcal{Q}_1(r_1), \dots, \mathcal{Q}_M(r_M), \mathcal{Q}_1(r_{M+1}), \dots,
  \mathcal{Q}_M(r_{MN}) \right]^T \; ,
 \end{aligned}
\label{Eq4a}
\end{equation}
where
\begin{equation}
\begin{aligned}
  &  {\bf r} = {\bf U}{\bf x}-{\bf V}{\bf y}, \\
 &  {\bf V} = {\bf I}_N \otimes \underbrace{\left[ \begin{array}{ccccc}
                            0 &0 &\dots &0 &0\\
                            e^{-j\psi} &0 &\dots &0 &0 \\
                            e^{-j2\psi} &e^{-j\psi} &\dots &0 &0 \\
                            &     &\vdots & & \\
                            e^{-j(M-1)\psi} &e^{-j(M-2)\psi}  & \dots &e^{-j\psi} &0
                           \end{array} \right]}_{{\bf V}_d},\\
& {\bf U} =  {\bf I}_N \otimes \underbrace{\left( {\bf I}_{M} + {\bf V}_d \right)}_{{\bf U}_d}.
\end{aligned}
\label{Eq4b}
\end{equation}
In the next section, we will present an equivalent linear model to describe the output of the $\Sigma\Delta$ array in terms of ${\bf r}$. This linear model, combined with the structure of the matrices ${\bf U}$ and ${\bf V}$, is used to formulate a recursion for the computation of the output power, a quantity that is essential for LMMSE channel estimation.

\section{Equivalent Linear Model \\ for the Spatial $\Sigma\Delta$ Array}\label{LinMod}

As in the scalar case described by~\eqref{equivlin}, we will represent the operation of the $\Sigma\Delta$ array using an equivalent linear model defined by
\begin{equation}
 \begin{aligned}
  {\bf y} = \boldsymbol{\Gamma} {\bf r} + {\bf q}
  \end{aligned}
 \label{Eq6}
\end{equation}
for a given matrix $\boldsymbol{\Gamma}$, where ${\bf q}$ is the equivalent quantization noise. There are an infinite number of such models, one for every choice of $\boldsymbol{\Gamma}$, and each will result in a particular definition for the quantization noise with differing statistical properties. One possible choice for the matrix $\boldsymbol{\Gamma}$ is obtained by making ${\bf r}$ uncorrelated with ${\bf q}$, i.e. $\mathbb{E}\left[ {\bf r} {\bf q}^H \right] =0$. This leads to $\boldsymbol{\Gamma} = {\bf C}_{yr} {\bf C}_r^{-1}$, where ${\bf C}_{yr} = \mathbb{E}\left[ {\bf y}{\bf r}^H\right]$ is the cross-covariance matrix between ${\bf y}$ and ${\bf r}$, and ${\bf C}_r $ is the covariance matrix of ${\bf r}$. Assuming the elements of ${\bf r}$ are jointly Gaussian, the Bussgang theorem~\cite{bussgang1952crosscorrelation} can be applied. This approach was used to derive channel estimates in~\cite{Li_channel} for conventional one-bit quantization, and in our initial work on the $\Sigma\Delta$ array~\cite{rao2019massive}\footnote{In line with recent literature, we distinguish the Bussgang theorem, which is used to compute the cross-correlations between the Gaussian input and Gaussian output of a nonlinear device, from the Bussgang decomposition which is used equivalently as a linear decomposition even when the input is not necessarily Gaussian.}. 
However, the resulting definition for the quantization noise does not have a physical interpretation in the context of Fig.~\ref{fig:fig1}.

Instead, we define $\boldsymbol{\Gamma}$ following the approach of~\cite{pirzadeh2019spectral}, applying the Bussgang decomposition element-wise such that $\mathbb{E}\left[r_m q_m^* \right] = 0$. With this definition, $\boldsymbol{\Gamma}$ becomes a diagonal matrix whose $m$th diagonal element, $\gamma_m$, is given by
\begin{equation}
\begin{aligned}
\gamma_m = \frac{\mathbb{E}\left[ r_m y_m^* \right]}{\mathbb{E}\left[| r_m|^2 \right]}.
\end{aligned}
    \label{Eq7}
\end{equation}
Plugging ${\bf r} = {\bf U}{\bf x} - {\bf V}{\bf y}$ into~(\ref{Eq6}), we get
\begin{equation}
\begin{aligned}
{\bf y} = \left({\bf I} + \boldsymbol{\Gamma} {\bf V} \right)^{-1} \boldsymbol{\Gamma}{\bf U} {\bf x} + \left({\bf I} + \boldsymbol{\Gamma} {\bf V} \right)^{-1} {\bf q}.
\end{aligned}
    \label{Eq8}
\end{equation}

The specific numerical value for $\gamma_m$ will depend on the output level $\alpha_{m'}$. We will choose $\alpha_{m'}$ such that $\gamma_m=1$, or equivalently such that $\boldsymbol{\Gamma} = {\bf I}_{MN}$, and~(\ref{Eq8}) can be simplified to \begin{equation}
    {\bf y} = {\bf x} + {\bf U}^{-1} {\bf q} \; .
    \label{Eq9}
\end{equation}
This {\color{black} is a convenient choice since the} model is now the exact spatial analog of the temporal $\Sigma\Delta$ architecture, and is equivalent to passing ${\bf x}$ through a (spatial) all-pass filter and ${\bf q}$ through a filter that shapes the quantization noise away from the AoA corresponding to $\psi$. {\color{black} While this choice for $\alpha_{m'}$ is an elegant one and provides good performance in our simulation results, we do not have a proof of its optimality. Such a proof appears to be difficult to obtain, so we leave it for future work. Nevertheless, a choice of $\gamma$ different from 1 seems difficult to justify since it would imply that the signal $\mathbf{x}$ is spatially filtered by the ADC architecture in some unclear way. As shown in the following sections for one- and two-bit quantization, the value of $\alpha_{m'}$ depends in general on the statistics of} the signal $r_m$, which would have to remain time-invariant for the quantizer gains to be fixed. This can be achieved in practice using an automatic gain control at the input to the ADC. 

\subsection{One-Bit Spatial $\Sigma\Delta$ ADCs}
When the $\Sigma\Delta$ array is implemented with one-bit ADCs, the output is given by
\begin{equation}
 y_m = \alpha_{m'} \left( {\rm sign}({\rm Re}(r_m)) + j {\rm sign}({\rm Im}(r_m)) \right),
 \label{Eq9a}
\end{equation}
and~(\ref{Eq7}) can be simplified to 
\begin{equation}
    \begin{aligned}
     \gamma_m &= \alpha_{m'} \frac{\mathbb{E}\left[|{\rm Re}(r_m)| +|{\rm Im}(r_m)| \right]}{\mathbb{E}\left[| r_m|^2 \right]} \\
& = \alpha_{m'} \frac{\mathbb{E}\left[ |{\rm Re}(r_m)| \right]}{\mathbb{E}\left[ |{\rm Re}(r_m)|^2 \right]}
    \end{aligned}
\end{equation}
for circularly symmetric $r_m$. Since the power of the pilot symbols is time-invariant, the statistics of $r_m$ are identical to those of $r_{m'}$. Consequently, we can set $\gamma_m=1$, which leads to
\begin{equation}
\begin{aligned}
\alpha_{m'} = \frac{\mathbb{E}\left[ |{\rm Re}(r_m)|^2 \right]}{\mathbb{E}\left[ |{\rm Re}(r_m)| \right]}.
\end{aligned}\label{alphamm}
\end{equation}

{\color{black} If $r_m$ were Gaussian, \eqref{alphamm} could be simplified to} 
\begin{equation}
    \alpha_{m'} = \frac{\sqrt{\pi \mathbb{E}\left[ |r_m|^2 \right] }}{2} \; .
    \label{Eq10}
\end{equation}
{\color{black} This was the value for $\alpha_{m'}$ used in~\cite{pirzadeh2019spectral}, and it provides sufficiently accurate estimates of the spectral efficiency for the case where the CSI is already known. However, the deviation of $r_m$ from Gaussianity, while not large, is sufficient to render~\eqref{Eq10} unsuitable for channel estimation.} The derivation of~\eqref{Eq10} relies on the fact that
\begin{equation}
\label{pi2}
    \frac{\sqrt{\mathbb{E}\left[ |{\rm Re}(r_m)|^2 \right] }}{\mathbb{E}\left[ |{\rm Re}(r_m)| \right]} = \sqrt{\frac{\pi}{2}}
\end{equation}
for Gaussian random variables. However, due to the non-linear feedback of the $\Sigma\Delta$ array, the tails of the distribution of $r_m$ are slightly heavier than a Gaussian, so the ratio on the left hand side of~\eqref{pi2} is slightly greater than $\sqrt{\pi/2}$. 

{\color{black} Whether Eq.~(\ref{alphamm}) or~\eqref{Eq10} is used to calculate $\alpha_{m'}$, in a practical implementation some empirical measurement of the mean power and absolute value of $r_m$ in the $\Sigma\Delta$ architecture would be necessary, and could be facilitated by the use of an automatic gain control. Rather than implementing the computation of $\alpha_{m'}$ according to~\eqref{alphamm}, in our simulations of the one-bit case presented later, we simply calculate $\alpha_{m'}$ as
\begin{equation}
\alpha_{m'}^* = \beta \frac{\sqrt{\pi \mathbb{E}\left[ |r_m|^2\right] }}{2},
    \label{Eq11}
\end{equation}
with a value of $\beta > 1$. There 
is a very small range of values near one that are appropriate for $\beta$. To see this,}
let  $\sigma_{r_m}^2 \triangleq \mathbb{E}\left[ |r_m|^2 \right]$. Using similar definitions for $\sigma_{y_m}^2$ and $\sigma_{q_m}^2$, and the fact that $r_m$ and $q_m$ are uncorrelated, we obtain the following relationship between the powers of the input and output of the array:
\begin{equation}
\begin{aligned}
\sigma_{y_m}^2 = \frac{\pi}{2} \beta^2 \sigma_{r_m}^2, \qquad
\sigma_{q_m}^2 = \sigma_{y_m}^2 - \sigma_{r_m}^2.
\end{aligned}
    \label{Eq12}
\end{equation}
From~(\ref{Eq12}), we can see that in order to prevent the quantization noise power from becoming greater than the input power $\sigma_{r_m}^2$, we must ensure that $\left(\frac{\pi}{2} \beta^2 -1 \right) < 1$, and hence that $1 \le \beta < 2/\sqrt{\pi} \approx 1.1284$. Otherwise, the input power to each ADC grows monotonically with the antenna index. In the simulations, we chose the value $\beta=1.05$.

\vspace*{-3mm}

\subsection{Two-bit Spatial $\Sigma\Delta$ ADCs}
\label{TwoBit}
{\color{black}
In this section, we extend the above analysis to the case where the quantizers in the $\Sigma\Delta$ array employ two-bits of resolution, implying four quantization levels.  Unlike the one-bit case, with two bits the Gaussian approximation for $r_m$ is quite accurate. We use the well-known Lloyd-max condition to determine the optimum quantization levels that minimize the distortion~\cite{max1960quantizing,lloyd1982least}. We will denote the quantization levels and the associated intervals that minimize the distortion for unit variance Gaussian inputs by $\nu_i$ and $\left( \nu_i^{\rm lo}, \nu_i^{\rm hi} \right), i=1,\dots,4$, respectively, and define
\begin{equation}
 \mathcal{Q}_{m'}\left( r_m^{\rm Re} \right) = \nu_i, \quad {\rm if} \,\, r_m^{\rm Re} \in \biggl( \biggr.  \frac{\sigma_{r_m}}{\sqrt{2}} \nu_i^{\rm lo},  \frac{\sigma_{r_m}}{\sqrt{2}} \nu_i^{\rm hi} \biggl. \biggr] \; ,
\end{equation}
where $ r_m^{\rm Re} \triangleq {\rm Re}(r_m)$. The above quantization levels satisfy $\nu_i^{\rm hi} = \nu_{i+1}^{\rm lo}$,  $\nu_1^{\rm lo} = -\infty$, and $\nu_4^{\rm hi} = \infty$, and the quantization bins have been adjusted to span the range of the input levels by modeling $r_m$ as a circularly symmetric Gaussian random variable with variance $\sigma_{r_m}^2$. Note that, while the convention is to also scale the output quantization level according to standard deviation of the input, we perform this scaling with the factor $\alpha_{m'}$ as we did for the one-bit case using~\eqref{Eq4}.

Assuming a linear model as before and using an element-wise Bussgang decomposition, (\ref{Eq7}) can be written as 
\begin{equation}
\begin{aligned}
    \gamma_m = \frac{\mathbb{E}\left[ r_m^{\rm Re}  y_m^{\rm Re}  \right]}{\mathbb{E}\left[  r_m^{\rm Re}  \right]^2}
    \label{Eq2b-2}
    \end{aligned}
\end{equation}
due to the circular symmetry of the data, where $y_m^{\rm Re} \triangleq {\rm Re} \left( y_m\right)$. The numerator of~\eqref{Eq2b-2} can be obtained from Bussgang's theorem~\cite{bussgang1952crosscorrelation}:
\begin{equation}
\begin{aligned}
  & \mathbb{E}\left[ r_m^{\rm Re}  y_m^{\rm Re}  \right] = \alpha_{m'} \mathbb{E} \left[r_m^{\rm Re} \mathcal{Q}_{m'} \left(r_m^{\rm Re} \right) \right] \\
   &= \alpha_{m'} \int_{-\infty}^{\infty} \frac{1}{\sqrt{2\pi}}  \frac{\partial \mathcal{Q}_{m'} \left(r_m^{\rm Re} \right) }{\partial r_m^{\rm Re} }  {\rm exp}\left(- \frac{\left(r_m^{\rm Re} \right)^2}{\sigma_{r_m}^2} \right) {\rm d}r_m^{\rm Re}.
   \end{aligned}
    \label{Eq2b-2a}
\end{equation}
The derivative $\partial \mathcal{Q}_{m'} \left(r_m^{\rm Re} \right) /\partial r_m^{\rm Re} $ can be computed as
\begin{equation}
   \frac{\partial \mathcal{Q}_{m'} \left(r_m^{\rm Re}\right) }{\partial r_m^{\rm Re}} =  \sum_{i=2}^{4} \left( \nu_i - \nu_{i-1} \right) \delta \left( r_m^{\rm Re}-  \frac{\sigma_{r_m}}{\sqrt{2}} \nu_i^{\rm lo} \right) \; ,
    \label{Eq2b-3}
\end{equation}
{\color{black} using the Dirac delta function $\delta(\cdot)$ to represent the derivative at the quantizer steps.}
Substituting the above equation in~(\ref{Eq2b-2a}) and evaluating the integral, we get 
\begin{equation}
   \begin{aligned}
&\mathbb{E}\left[ r_m^{\rm Re} y_m^{\rm Re} \right] = \alpha_{m'} \sum_{i=2}^{4} \frac{\left( \nu_i - \nu_{i-1} \right)}{\sqrt{2\pi}} {\rm exp} \left( - \frac{\left( \nu_i^{\rm lo}\right)^2}{2} \right).
  \end{aligned}
  \label{Eq2b-4}
\end{equation}
Then, the value of $\alpha_{m'}$ that yields $\gamma_m=1$ is given by 
\begin{equation}
   \begin{aligned}
\alpha_{m'} = \frac{\sigma_{r_m}\sqrt{\pi/2} }{  \sum_{i=2}^{4} \left( \nu_i - \nu_{i-1} \right) {\rm exp} \left( - \frac{\left( \nu_i^{\rm lo}\right)^2}{2} \right)}.
  \end{aligned}
  \label{Eq2b-4a}
\end{equation}
Thus, $\alpha_{m'}$ is determined by the standard deviation of the ADC input, the quantization intervals and the corresponding output levels. Finally, computing the expectation $\mathbb{E}\left[ | y_m |^2 \right] $, the output and quantization noise powers are, respectively, given by 
\begin{equation}
\begin{aligned}
\sigma_{y_m}^2 &= \alpha_{m'}^2 \sum_i \left| \mathcal{Q}_{m'}\left( r_m^{\rm Re} \right) \right|^2 {\rm Pr}\left(\frac{\sigma_{r_m}}{\sqrt{2}} \nu_i^{\rm lo} < r_m^{\rm Re} \leq  \frac{\sigma_{r_m}}{\sqrt{2}} \nu_i^{\rm hi}  \right) \\
& = 2 \alpha_{m'}^2 \sum_{i=1}^{4} \nu_i^2 \left( \Psi \left( \frac{\sigma_{r_m}}{\sqrt{2}} \nu_i^{\rm hi} \right) - \Psi \left( \frac{\sigma_{r_m}}{\sqrt{2}} \nu_i^{\rm lo} \right) \right), \\
\sigma_{q_m}^2 &= \sigma_{y_m}^2 - \sigma_{r_m}^2,
\end{aligned}
    \label{Eq2b-5}
\end{equation}
where $\Psi\left( \cdot \right)$ is the cumulative distribution function of the normal distribution. }

\section{Channel Estimation with spatial $\Sigma\Delta$ ADCs}
\label{ChEst}
By describing the input-output relationship with an equivalent linear model, we were able to arrive at closed-form expressions for the quantization noise and output powers in Eqs.~\eqref{Eq12} and~\eqref{Eq2b-5}. Leveraging the results obtained in the previous section, we derive below the LMMSE channel estimate based on the one-bit or two-bit outputs ${\bf y}$ of the $\Sigma\Delta$ ADC array. 
\subsection{LMMSE Channel Estimation} \label{sub:LMMSE}
The LMMSE channel estimate is defined by
\begin{equation}
\begin{aligned}
    \hat{\bf g} &= \mathbb{E}\left[ {\bf g}{\bf y}^H \right] \left( \mathbb{E}\left[ {\bf y}{\bf y}^H \right] \right)^{-1} {\bf y} \\
    &= {\bf C}_{gy} {\bf C}_y^{-1} {\bf y} \; .
    \end{aligned}
    \label{Eq18}
\end{equation}
Using~(\ref{Eq9}), we can obtain the covariance matrix of ${\bf y}$: 
\begin{equation}
\begin{aligned}
{\bf C}_y &= {\bf C}_x + {\bf U}^{-1} {\bf C}_q {\bf U}^{-H},
\end{aligned}
    \label{Eq13}
\end{equation}
where ${\bf C}_q$ is the covariance matrix of ${\bf q}$. {\color{black} The expression in~\eqref{Eq13} relies on the assumption that $\mathbb{E}\left[ {\bf x}{\bf q}^H \right] = {\bf 0}$, which we show in Appendix~A to be true if $r_m$ is Gaussian\footnote{{\color{black}This does not imply that $ \mathbb{E}\left[ {\bf r}{\bf q}^H \right] = {\bf 0}$, as ${\bf r}$ and ${\bf q}$ are clearly correlated.}}. Although $r_m$ is strictly speaking not Gaussian, the assumption is sufficiently accurate here, and we will see in Section~\ref{SimResults} that it yields a channel estimator with good performance.} Similarly, since we have chosen $\boldsymbol{\Gamma}={\bf I}$ in~(\ref{Eq6}), it is easy to show that
\begin{equation}
\begin{aligned}
{\bf r} =  {\bf x} - {\bf V}{\bf U}^{-1}{\bf q}
\end{aligned}
    \label{Eq15a}
\end{equation}
and hence that
\begin{equation}
\begin{aligned}
{\bf C}_r &= {\bf C}_x +  {\bf V}{\bf U}^{-1} {\bf C}_q {\bf U}^{-H}{\bf V}^H.
\end{aligned}
    \label{Eq15}
\end{equation}
From~(\ref{Eq13}), we see that ${\bf C}_y$ is determined by ${\bf C}_q$, whereas~(\ref{Eq12}) and~(\ref{Eq2b-5}) show that the quantization noise power is dependent on $\sigma_{y_m}^2$. Due to this inter-relationship between ${\bf C}_y$ and ${\bf C}_q$, these matrices cannot be computed in closed form. However, they can be computed in a recursive manner. 

Using $\mathbb{E}\left[ {\bf x}{\bf q}^H \right] \approx {\bf 0}$ and the fact that $\mathbb{E}\left[ r_m q_m^* \right] = 0$, we can show that $\mathbb{E}\left[ q_m q_{m \pm 1}^* \right] \approx 0$. As a result, ${\bf C}_q$ is approximately diagonal with elements given by $\sigma_{q_m}^2$. Furthermore, noting that ${\bf V}{\bf U}^{-1} $ has the structure 
\begin{equation}
 {\bf V} {\bf U}^{-1} = {\bf I}_N \otimes e^{-j\psi} \left[ \begin{array}{ccccc}
                            0 &0 &\dots &0 &0\\
                            1 &0 &\dots &0 &0 \\
                            0 &1 &\dots &0 &0 \\
                            &     &\vdots & & \\
                            0 &0  & \dots &1 &0
                           \end{array} \right],
    \label{Eq16}
\end{equation}
we can generate the following recursion for $\sigma_{r_m}^2$, $\sigma_{y_m}^2$ and $\sigma_{q_m}^2$ using~(\ref{Eq13}) and~(\ref{Eq15}):
\begin{equation}
\begin{aligned}
& \sigma_{r_m}^2 = \begin{cases}
\sigma_{x_m}^2, \quad m=kM+1,\quad  k= 0,1,\dots,N-1,\\
\\
\sigma_{x_m}^2 + \sigma_{q_{m-1}}^2, \quad \text{otherwise}.
\end{cases} \\
\\
& \sigma_{y_m}^2 = \begin{cases}
\frac{\pi}{2} \beta^2 \sigma_{r_m}^2, \quad \text{for 1-bit ADCs} \\
\\
\frac{\alpha^2 \sigma_{r_m}^2}{2} \sum_{i=1}^{4} \nu_i^2 \left( \Psi \left( \frac{\sigma_{r_m}}{\sqrt{2}} \nu_i^{\rm hi} \right) - \Psi \left( \frac{\sigma_{r_m}}{\sqrt{2}} \nu_i^{\rm lo} \right) \right), \\
 \hspace{1.75cm} \text{for 2-bit ADCs} \\
\end{cases} \\
\vspace{3em}
\\
& \sigma_{q_m}^2 = \sigma_{y_m}^2 - \sigma_{r_m}^2.
\end{aligned}
    \label{Eq17}
\end{equation}
In the above equations, $\sigma_{x_m}^2$ is the $m$th diagonal element of ${\bf C}_x$. Thus, the power of the $m$th quantizer input, $\sigma_{r_m}^2$, depends on the quantization noise powers computed up to index $m-1$. Then, the $m$th output power, $\sigma_{y_m}^2$, is given by~(\ref{Eq12}) for one-bit ADCs and by~(\ref{Eq2b-5}) for two-bit ADCs. This allows us to compute $\sigma_{q_m}^2$, and from there $\sigma_{r_{m+1}}^2$, and so on. Following this process for indices $m = 1$ to $MN$ allows us to compute ${\bf C}_q$, and finally the complete ${\bf C}_y$ is obtained from~(\ref{Eq13}). 

Thus, we can obtain the LMMSE estimate of the channel, $\hat{\bf g}$, from~(\ref{Eq18}) where 
\begin{equation}
\begin{aligned}
   {\bf C}_{gy} &= \mathbb{E}\left[ {\bf g}{\bf y}^H \right] \\
   &= \mathbb{E}\left[ {\bf g} \left(\boldsymbol{\Phi} {\bf g} + {\bf n}+ {\bf U}^{-1} {\bf q}\right)^H \right] \\
   & = {\bf C}_g \boldsymbol{\Phi}^H + \mathbb{E}\left[ {\bf g} {\bf q}^H \right]{\bf U}^{-H} \\
   & \approx {\bf C}_g \boldsymbol{\Phi}^H \; .
  \end{aligned}
  \label{Eq18-a}
\end{equation}
The final approximation results because $\mathbb{E} \left[ {\bf g} {\bf q}^H \right] = \boldsymbol{\Phi}^{\dagger} \mathbb{E} \left[ {\bf x} {\bf q}^H \right]- \boldsymbol{\Phi}^{\dagger} \mathbb{E}\left[{\bf n} {\bf q}^H \right] \approx {\bf 0}$, since $\mathbb{E} \left[ {\bf x} {\bf q}^H \right] \approx {\bf 0}$ and we can show that $\mathbb{E} \left[ {\bf n} {\bf q}^H \right] = {\bf 0}$ using an argument identical to that in Appendix A for Gaussian noise ${\bf n}$.
The resulting algorithm for computing the LMMSE channel estimate for the $\Sigma\Delta$ array has low complexity and is summarized in Algorithm~1.

\begin{algorithm}
\SetAlgoLined
\begin{enumerate}[wide, labelwidth=!, labelindent=0pt]
\item Set $\beta = 1.05$ for one-bit operation. For $m= 1$ to $MN$, repeat:
\begin{enumerate}[label=(\roman*),wide, labelwidth=!, labelindent=0pt]
 \item Update the diagonal elements of ${\bf C}_r$, $\sigma_{r_m}^2$, using~(\ref{Eq17}). Update $\sigma_{y_m}^2$ using~(\ref{Eq12}) for one-bit ADCs and~(\ref{Eq2b-5}) for two-bit ADCs. 
 \item The elements of ${\bf C}_q$, $\sigma_{q_m}^2$, are updated using $\sigma_{q_m}^2 = \sigma_{y_m}^2 - \sigma_{r_m}^2$. 
 \item Update $\alpha_{m'} = \begin{cases}
\text{From~(\ref{Eq11}) for one-bit $\Sigma\Delta$ ADCs}, 
\\
\text{From~(\ref{Eq2b-4a}) for two-bit $\Sigma\Delta$ ADCs}.
\end{cases}.
$
\end{enumerate}
\item Obtain the complete matrix ${\bf C}_y$ using~(\ref{Eq13}).
\item Compute the output of the $\Sigma\Delta$ array as follows:
\begin{enumerate}[label=(\roman*),wide, labelwidth=!, labelindent=0pt]
 \item $r_m = x_m$, for $m=kM+1, k=0,\dots,N-1$.\\
\quad\, $r_m = x_m + e^{-j \psi} \left( r_{m-1} - y_{m-1} \right)$, otherwise.\\
\item $y_m = \alpha_{m'} \left( \mathcal{Q}_m \left( {\rm Re}(r_m) \right) + j \mathcal{Q}_m \left( {\rm Im}(r_m) \right)  \right)$.
\end{enumerate}
\item Estimate the channel, $\hat{\bf g}$, from~(\ref{Eq18}).
\end{enumerate}
  \caption{Channel estimation using $\Sigma\Delta$ array}
\end{algorithm}

\subsection{Channel Estimation Error}
{\color{black} As we will be comparing the $\Sigma\Delta$ channel estimation error with standard one-bit sampling, which is unable to identify the channel gain, we will use normalized error (or cosine distance)} that is independent of any scaling factor affecting either the real channel or its estimate:
\begin{equation}
\begin{aligned}
{\rm NE} &=\min_{\zeta} \frac{\mathbb{E}\left[\lVert{\bf g}-\zeta\hat{\bf g}\rVert_2^2\right]}{\mathbb{E}\left[\lVert{\bf g}\rVert_2^2\right]}\\
&=1-\frac{\mathbb{E}\left[\lvert\hat{\bf g}^H{\bf g}\rvert^2\right]}{\mathbb{E}\left[\lVert\hat{\bf g}\rVert_2^2\right]\mathbb{E}\left[\lVert{\bf g}\rVert_2^2\right]}.
\end{aligned}
\label{NMSE_def}
\end{equation}
This expression for the NE, {\color{black} which satisfies $0 \le \mbox{\rm NE} \le 1$,} is valid for any estimator. For the particular case of LMMSE estimators, which necessarily satisfy $\mathbb{E}\left[{\bf g}\hat{\bf g}^H\right]={\bf C}_{\hat{g}}$, (\ref{NMSE_def}) can be expressed as
\begin{equation}
{\rm NE} = \frac{{\mathbb{E}\left[ \lVert {\bf g} \rVert_2^2 \right]}- {\mathbb{E}\left[ \lVert \hat{\bf g} \rVert_2^2 \right]}}{\mathbb{E}\left[ \lVert {\bf g} \rVert_2^2 \right]} = \frac{{\rm Tr\left({\bf C}_{g}-{\bf C}_{\hat{g}} \right)}}{{\rm Tr}\left( {\bf C}_{g} \right)},
\label{Eq20}
\end{equation}
where ${\bf C}_{\hat{g}}$ is given by \begin{equation}
\begin{aligned}
    {\bf C}_{\hat{g}} &= \mathbb{E}\left[ \hat{\bf g} \hat{\bf g}^H \right] = {\bf C}_g \boldsymbol{\Phi} ^H {\bf C}_y^{-1} \boldsymbol{\Phi} {\bf C}_g \\ 
    &={\bf C}_g \boldsymbol{\Phi} \left( {\bf C}_x +  {\bf U}^{-1} {\bf C}_q {\bf U}^{-H} \right)^{-1} \boldsymbol{\Phi}^H {\bf C}_g.
    \label{Eq19}
    \end{aligned}
\end{equation}
Finally, the estimation error covariance matrix is given by
\begin{equation}
{\bf C}_{\epsilon} = {\bf C}_g - {\bf C}_g \boldsymbol{\Phi} \left( {\bf C}_x +  {\bf U}^{-1} {\bf C}_q {\bf U}^{-H} \right)^{-1} \boldsymbol{\Phi}^H {\bf C}_g.
    \label{Eq21}
\end{equation}
When the pilots are orthogonal and $N=K$, the matrices $ {\bf C}_x$, $ {\bf C}_y$ and $ {\bf C}_q$ are block-diagonal with identical blocks. {\color{black}As a result, ${\bf C}_{\hat{g}}$ is also block-diagonal and the $k$th $M \times M$ block corresponds to the covariance matrix of the estimated channel of the $k$th user.} 


\section{Uplink Achievable Rate Analysis}
\label{UplinkRate}
In this section, we derive the uplink achievable rate for MRC and ZF receivers. In the uplink data transmission stage, the $K$ users transmit their data represented by the $K\times 1$ vector~${\bf s}$. Using a Bussgang decomposition as described previously on the $\Sigma\Delta$-quantized received signal, ${\bf y}_d$, we get 
\begin{equation}
\begin{aligned}    
{\bf y}_d =& \mathcal{Q}\left({\bf r}_d \right) =\sqrt{\rho_d} {\bf G} {\bf s} + {\bf n}_d + {\bf U}_d^{-1} {\bf q}_d,
    \end{aligned}
    \label{Eq22}
\end{equation}
where ${\bf r}_d = {\bf U}_d \left( \sqrt{\rho_d} {\bf G} {\bf s} + {\bf n}_d \right) - {\bf V}_d {\bf y}_d$, $\rho_d$ is the data transmission power, ${\bf n}_d \sim \mathcal{CN}\left({\bf 0}, {\bf C}_N \right)$ and ${\bf q}_d$ are the additive and quantization noise in the data phase, respectively. The matrices ${\bf U}_d$ and ${\bf V}_d$ are defined by taking $N=1$ in Eq.~\eqref{Eq4b}. {\color{black}For this analysis, we assume that the user symbols $s_k$ are i.i.d with $\mathbb{E}\left[ |s_k|^2 \right]=1$ and that the channel covariance matrices of the different users are equal and denoted by ${\bf C}_G={\bf C}_{g_1}=\cdots={\bf C}_{g_K}$.} 

The analysis of the achievable rate relies on the covariance matrix of ${\bf q}_d$, ${\bf C}_{q_d}$, which is different from the quantization noise covariance matrix during the training phase. For the data transmission stage, this matrix has to be derived in a manner similar to Section~\ref{ChEst}. Inspecting the recursion equations developed in the previous section, we see that initialization of the recursion will be performed with ${\bf C}_{x_d}$, where ${\bf C}_{x_d} = \rho_d {\bf G} {\bf G}^H + {\bf C}_N$. To simplify the subsequent analysis, as in \cite{Li_channel,mollen217tpc} we approximate
${\bf G} {\bf G}^H$ by {\color{black}$K{\bf C}_G$, which becomes increasingly accurate as $K$ grows. We will see in the simulations that excellent agreement between the theoretical and simulated spectral efficiency is obtained even for values as low as $K=10$.}
Thus
\begin{equation}
{\bf C}_{x_d} \approx K \rho_d {\bf C}_G + {\bf C}_N.
\label{Eq22a}    
\end{equation}

The procedure to obtain ${\bf C}_{q_d}$ is outlined as follows. Let $\sigma_{r_{d_m}}^2$, $\sigma_{y_{d_m}}^2$ and $\sigma_{q_{d_m}}^2$ denote the powers of the $m$th components of ${\bf r}_d$, ${\bf y}_d$ and ${\bf q}_d$, respectively. Then,~(\ref{Eq17}) is modified for the data transmission stage as
\begin{equation}
\begin{aligned}
& \sigma_{r_{d_m}}^2 = \begin{cases}
\sigma_{x_{d_m}}^2, \quad m=0,1,\dots,M, \\
\\
\sigma_{x_{d_m}}^2 + \sigma_{q_{d_{m-1}}}^2, \quad \text{otherwise}.
\end{cases} \\
\\
& \sigma_{y_{d_m}}^2 = \begin{cases}
\frac{\pi}{2} \beta^2 \sigma_{r_{d_m}}^2, \quad \text{for 1-bit ADCs} \\
\\
\frac{\alpha^2 \sigma_{r_{d_m}}^2}{2} \sum_{i=1}^{4} \nu_i^2 \left( \Psi \left( \frac{\sigma_{r_{d_m}}}{\sqrt{2}} \nu_i^{\rm hi} \right) - \Psi \left( \frac{\sigma_{r_{d_m}}}{\sqrt{2}} \nu_i^{\rm lo} \right) \right), \\
 \hspace{1.75cm} \text{for 2-bit ADCs} \\
\end{cases} \\
\vspace{3em}
\\
& \sigma_{q_{d_m}}^2 = \sigma_{y_{d_m}}^2 - \sigma_{r_{d_m}}^2,
\end{aligned}
    \label{Eq22b}
\end{equation}
where $\sigma_{x_{d_m}}^2$ is the $m$th diagonal element of ${\bf C}_{x_d}$. The diagonal matrix $ {\bf C}_{q_d}$ is completed with $\sigma_{q_{d_m}}^2$ as its diagonal elements.

The BS uses a linear receiver for symbol detection that depends on the LMMSE channel estimate. Denoting the linear receiver by ${\bf W}$, the detected symbol vector is obtained by multiplying the conjugate transpose of ${\bf W}$ with the received signal vector as
\begin{equation}
 \hat{\bf s} = {\bf W}^H {\bf y}_d = \sqrt{\rho_d} {\bf W}^H  {\bf G} {\bf s} +  {\bf W}^H{\bf n}_d + {\bf W}^H{\bf U}_d^{-1} {\bf q}_d.
    \label{Eq23}
\end{equation}
From~(\ref{Eq23}), we can re-write the various components contributing to the $k$th detected symbol as
\begin{equation}
\begin{aligned}
    \hat{s}_k = &\sqrt{\rho_d} \mathbb{E} \left[ {\bf w}_k^H {\bf g}_k \right] s_k + \sqrt{\rho_d} \left( {\bf w}_k^H {\bf g}_k  - \mathbb{E} \left[ {\bf w}_k^H {\bf g}_k \right] \right) s_k + \\ 
    & \sqrt{\rho_d} {\bf w}_k^H \sum_{i \neq k} {\bf g}_i s_i +  {\bf w}_k^H {\bf n}_d +  {\bf w}_k^H {\bf U}_d^{-1} {\bf q}_d,
    \end{aligned}
\label{Eq24}
\end{equation}
where ${\bf w}_k$ is the $k$th column of ${\bf W}$. The terms in the above equation correspond to the desired signal, the receiver uncertainty, the inter-user interference, the additive noise and the quantization noise, respectively. 

We will follow an approach similar to~\cite{mollen217tpc} and~\cite{zhang2014power} and use the classical worst-case uncorrelated Gaussian assumption on the terms in~(\ref{Eq24}) to obtain a lower bound on the achievable rate. Treating the final four terms as ``effective noise", the achievable rate of the $k$th user is 
given by~(\ref{Eq25}) at the top of the next page, based on the widely used approximation for massive MIMO systems~\cite{zhang2014power}
\begin{equation}
    \mathbb{E}\left[{\rm log}_2\left( 1 + \frac{X}{Y} \right) \right]\approx {\rm log}_2 \left(1 + \frac{\mathbb{E}[X]}{\mathbb{E}[Y]}\right),
    \label{Eq24a}
\end{equation}
where $X$ and $Y$ are sums of non-negative random variables. {\color{black} The above approximation becomes increasingly accurate for a large number of antennas since, according to the law of large numbers, the variances of both X and Y become small due to channel hardening effect.} Whereas the achievable rate bounds derived in~\cite{pirzadeh2019spectral} assume perfect knowledge of the CSI, our result takes into account the channel estimation error. In our derivation of the worst-case bound, we assume that the channel estimate $\hat{\bf g}$ is Gaussian with covariance matrix $ {\bf C}_{\hat{g}}$ given by~(\ref{Eq19}). Similarly, ${\bf q}_d$ is also assumed to be Gaussian and its covariance matrix is obtained as described earlier in this section. In the following, we consider the performance for the specific cases of the MRC, ZF and LMMSE receivers. In the derivation of the achievable rate using these receivers, we will assume that during the training phase, the pilots are orthogonal and that $N=K$. Consequently, the matrices ${\bf C}_x$, ${\bf C}_y$ and ${\bf C}_{\hat{g}}$ will be block-diagonal.
\color{black}
 \begin{figure*}[htb]
{\begin{equation}
\begin{aligned}
 R_k = {\rm log}_2\left( 1 + \frac{\rho_d \left\lvert \mathbb{E} \left[ {\bf w}_k^H {\bf g}_k \right] \right\rvert^2 }{\rho_d {\rm var} \left( {\bf w}_k^H {\bf g}_k \right)  + \rho_d \sum_{i \neq k} \mathbb{E}\left[ |{\bf w}_k^H {\bf g}_i|^2 \right] + \mathbb{E}\left[ \left\lvert {\bf w}_k^H {\bf n}_d \right\rvert^2 \right] + \mathbb{E} \left[ \left\lvert {\bf w}_k^H {\bf U}_d^{-1} {\bf q}_d \right\rvert^2 \right]  } \right)
\end{aligned}
\label{Eq25}
\end{equation}}
\end{figure*}

\subsection{MRC receiver}
To simplify the analysis, we consider an MRC receiver without pre-whitening, as follows:
\begin{equation}
\begin{aligned}
&{\bf W}_{\rm MRC} = \hat{\bf G},
 \end{aligned}    
   \label{Eq23a1}
\end{equation}
where $\hat{\bf G}$ is the $M \times K$ matrix formed from $\hat{\bf g}$ using the inverse of the ${\rm vec}$ operation. The achievable rate of the $k$th user is given by~(\ref{Eq26}) on the next page. 
\begin{figure*}[htb]
\begin{equation}
\begin{aligned}
 R_k^{\rm MRC} = {\rm log}_2\left( 1 + \frac{\rho_d {\rm Tr} \left( {\bf C}_{\hat{g}}\right)  /K }{\rho_d \, K \, {\rm Tr} \left( {\bf C}_{G}\right) +  {\rm Tr}\left({\bf C}_N \right) + {\rm Tr}\left( {\bf U}_d^{-1} {\bf C}_{q_d}  {\bf U}_d^{-H} \right)  } \right)
\end{aligned}
\label{Eq26}
\end{equation}
\end{figure*}
To show this, we compute below the individual terms in the achievable rate expression of~(\ref{Eq25}).
\begin{equation*}
\begin{aligned}
    \mathbb{E}\left[ \hat{\bf g}_k^H {\bf g}_k\right] =&  \mathbb{E}\left[ \hat{\bf g}_k^H \left(\hat{\bf g}_k + \boldsymbol{\epsilon}_k \right) \right] \\
    =& \mathbb{E} \left[ \lVert \hat{\bf g}_k \rVert^2  \right] + \mathbb{E} \left[  \hat{\bf g}_k^H \boldsymbol{\epsilon}_k  \right] \\
    = & \mathbb{E} \left[ \lVert \hat{\bf g}_k \rVert^2 \right] = {\rm Tr} \left({\bf C}_{\hat{g}_k} \right) = \frac{{\rm Tr} \left({\bf C}_{\hat{g}} \right)}{K},
    \end{aligned}
    \label{EqApp1}
\end{equation*}
where we have used the fact that the LMMSE channel estimate is uncorrelated with the channel estimation error and that the covariance matrices of the estimated channels for each of the users are equal. Further,
\begin{equation*}
\begin{aligned}
   & {\rm var}\left( \hat{\bf g}_k^H {\bf g}_k\right) \\
   =&  \mathbb{E}\left[ \left \lvert \hat{\bf g}_k^H {\bf g}_k \right \rvert^2 \right] - \left(\mathbb{E}\left[ \hat{\bf g}_k^H {\bf g}_k\right] \right)^2\\
    =& \mathbb{E} \left[ \left \lvert \lVert \hat{\bf g}_k \rVert^2  + \hat{\bf g}_k^H \boldsymbol{\epsilon}_k \right \rvert^2  \right] - \left(\mathbb{E}\left[ \hat{\bf g}_k^H {\bf g}_k\right] \right)^2 \\
    =& \mathbb{E} \left[\lVert \hat{\bf g}_k  \rVert^4 \right] +  \mathbb{E} \left[\lVert \hat{\bf g}_k  \rVert^2 \hat{\bf g}_k^H \boldsymbol{\epsilon}_k  \right] + \mathbb{E} \left[\lVert \hat{\bf g}_k  \rVert^2 \boldsymbol{\epsilon}_k^H \hat{\bf g}_k \right] +\\
    & \quad  \mathbb{E} \left[\lVert \hat{\bf g}_k^H \boldsymbol{\epsilon}_k  \rVert^2 \right] -  \left(\mathbb{E}\left[ \hat{\bf g}_k^H {\bf g}_k\right] \right)^2\\
    =& 2 \left(  \mathbb{E} \left[ \lVert \hat{\bf g}_k \rVert^2 \right] \right)^2 +  \mathbb{E} \left[ \lVert \hat{\bf g}_k \rVert^2 \right]  \mathbb{E} \left[ \lVert \boldsymbol{\epsilon}_k \rVert^2 \right] -  \left(\mathbb{E}\left[ \hat{\bf g}_k^H {\bf g}_k\right] \right)^2 \\
    =& \left( {\rm Tr} \left( {\bf C}_{\hat{g}_k} \right) \right)^2 +  {\rm Tr} \left( {\bf C}_{\hat{g}_k}\right) \left[  {\rm Tr} \left( {\bf C}_{G}\right) -  {\rm Tr} \left( {\bf C}_{\hat{g}_k}\right) \right] \\
    =& {\rm Tr} \left( {\bf C}_{\hat{g}_k}\right) {\rm Tr} \left( {\bf C}_{G}\right) = \frac{{\rm Tr} \left( {\bf C}_{\hat{g}}\right) {\rm Tr} \left( {\bf C}_{G}\right)}{K},
    \end{aligned}
    \label{EqApp2}
\end{equation*}
and, for $i \neq k$,
\begin{equation*}
\begin{aligned}
 &\mathbb{E}\left[ \left \lvert \hat{\bf g}_k^H {\bf g}_i \right \rvert^2 \right] = \left \lvert \mathbb{E} \left[ \hat{\bf g}_k^H {\bf g}_i  \right] \right \rvert^2 + \mathbb{E} \left[ \lVert \hat{\bf g}_k\rVert^2 \right] \mathbb{E} \left[ \lVert {\bf g}_i \rVert^2 \right],  \\
&\mathbb{E} \left[ \hat{\bf g}_k^H {\bf g}_i  \right] =  \mathbb{E} \left[ \left( {\bf P}_k \boldsymbol{\Phi} {\bf g} + {\bf P}_k  {\bf n} +  {\bf P}_k {\bf U}^{-1}  {\bf q}\right)^H {\bf g}_i \right],
    \end{aligned}
    \label{EqApp3}
\end{equation*}
where ${\bf P}_k = {\bf C}_{g_{(k-1)M:kM,:}} \boldsymbol{\Phi} {\bf C}_y^{-1} $ and ${\bf C}_{g_{(k-1)M:kM,:}}$ refers to the $k$th block of rows of ${\bf C}_{g}$, and where we have substituted the expression for $\hat{\bf g}_k$ in the above equation. Then, using the fact that ${\bf g}_i$ is uncorrelated with the quantization noise and the additive noise, we get
\begin{equation*}
\begin{aligned}
 &\mathbb{E}\left[ \left \lvert \hat{\bf g}_k^H {\bf g}_i \right \rvert^2 \right] = \frac{{\rm Tr} \left( {\bf C}_{\hat{g}}\right)  {\rm Tr} \left( {\bf C}_{G}  \right)}{K}.
    \end{aligned}
    \label{EqApp5}
\end{equation*}
We can solve for the final term in the denominator in a similar manner to get:
\begin{equation*}
\begin{aligned}
 &\mathbb{E}\left[ \left| \hat{\bf g}_k^H {\bf U}_d^{-1} {\bf q} \right|^2 \right] =  \frac{{\rm Tr}\left({\bf C}_{\hat{g}}\right) {\rm Tr}\left( {\bf U}_d^{-1} {\bf C}_{q_d}  {\bf U}_d^{-H} \right)}{K}. 
    \end{aligned}
    \label{EqApp6}
\end{equation*}
Note that when there is perfect knowledge of the CSI, i.e. ${\bf w}_k = {\bf g}_k$,~(\ref{Eq26}) reduces to the expression derived in~\cite{pirzadeh2019spectral}. The ratio inside the logarithm of~\eqref{Eq26} shows the impact of the various system parameters on the SNR for each of the users. The three terms in the denominator show the contribution of the multi-user interference (MUI), the receiver noise, and the quantization noise, respectively.
\vspace*{-4mm}
\subsection{ZF receiver}
The composite noise at the output of the $\Sigma\Delta$ array is spatially correlated and its covariance matrix is given by ${\bf C}_{\tilde{n}} = \left({\bf C}_N + {\bf U}_d^{-1} {\bf C}_{q_d} {\bf U}_d^{-H} \right)$. Thus, the ZF equalizer becomes
\begin{equation}
{\bf W}_{\rm ZF} = {\bf C}_{\tilde{n}}^{-1} \hat{\bf G} \left(\hat{\bf G}^H  {\bf C}_{\tilde{n}}^{-1} \hat{\bf G} \right)^{-1} \; .
    \label{Eq23a2}
\end{equation}
The expectations required to compute~\eqref{Eq23a2} are significantly more complicated than for the case of MRC, and are intractable to evaluate in closed-form. The achievable rate of the $k$th user can be expressed as in~(\ref{Eq27}), where the term $\mathbb{E}\left[ \left( \hat{\bf G}^H  {\bf C}_{\tilde{n}}^{-1} \hat{\bf G} \right)_{kk}\right]$ is computed empirically.
The remaining terms in the expression can be found as follows. For the numerator,
\begin{equation}
    \mathbb{E}\left[ {\bf w}_k^H {\bf g}_k \right] = \mathbb{E}\left[ {\bf w}_k^H \left( \hat{\bf g}_k + \boldsymbol{\epsilon}_k \right) \right] =1 + \mathbb{E}\left[ {\bf w}_k^H \boldsymbol{\epsilon}_k \right]  = 1 \; .
\end{equation}
Further,
\begin{equation*}
    {\rm var} \left( {\bf w}_k^H {\bf g}_k \right) = \mathbb{E}\left[ \left\lVert {\bf w}_k^H \boldsymbol{\epsilon}_k \right\rVert^2 \right] = \frac{ \mathbb{E}\left[ \left( \hat{\bf G}^H {\bf C}_{\tilde{n}}^{-1}\hat{\bf G} \right)_{kk}\right] {\rm Tr}\left( {\bf C}_{\epsilon} \right) }{K}.
\end{equation*}
For $i \neq k$, we have
\begin{equation*}
   \mathbb{E}\left[ \left \lvert {\bf w}_k^H {\bf g}_i \right \rvert^2 \right] =  \mathbb{E}\left[ \left \lvert {\bf w}_k^H \boldsymbol{\epsilon}_i \right \rvert^2 \right] = \frac{ \mathbb{E}\left[ \left( \hat{\bf G}^H {\bf C}_{\tilde{n}}^{-1} \hat{\bf G} \right)_{kk}\right] {\rm Tr}\left( {\bf C}_{\epsilon} \right) }{K}.
\end{equation*}
Similarly, we obtain
\begin{equation*}
\begin{aligned}
& \mathbb{E}\left[ \left| {\bf w}_k^H {\bf n}_d \right|^2 \right] =  \mathbb{E}\left[ \left( \hat{\bf G}^H {\bf C}_{\tilde{n}}^{-1} \hat{\bf G} \right)_{kk}^{-1}\right] {\rm Tr}\left({\bf C}_N \right),\\
   & \mathbb{E}\left[ \left| {\bf w}_k^H {\bf U}_d^{-1} {\bf q} \right|^2 \right] = \mathbb{E}\left[ \left( \hat{\bf G}^H {\bf C}_{\tilde{n}}^{-1} \hat{\bf G} \right)_{kk}^{-1}\right] {\rm Tr}\left( {\bf U}_d^{-1} {\bf C}_{q_d}  {\bf U}_d^{-H} \right) .
    \end{aligned}
\end{equation*}
\begin{figure*}[htb]
\begin{equation}
\begin{aligned}
 R_k^{\rm ZF} = {\rm log}_2\left( 1 + \frac{\rho_d }{\rho_d    \mathbb{E}\left[ \left( \hat{\bf G}^H  {\bf C}_{\tilde{n}}^{-1} \hat{\bf G} \right)_{kk}^{-1}\right] {\rm Tr}\left( {\bf C}_{\epsilon} \right)  +  \mathbb{E}\left[ \left( \hat{\bf G}^H  {\bf C}_{\tilde{n}}^{-1} \hat{\bf G} \right)_{kk}^{-1}\right]  {\rm Tr}\left( {\bf C}_{\tilde{n}} \right) } \right)
\end{aligned}
\label{Eq27}
\end{equation}
\end{figure*}
\vspace*{-4mm}
\subsection{LMMSE receiver}
The LMMSE receiver minimizes the mean squared error in estimating $s_k$ and can be expressed as~\cite{hien_mmse,li2015multi}
\begin{equation}
{\bf W}_{\rm MMSE} =   \left(\rho_d  \hat{\bf G} \hat{\bf G}^H + \rho_d {\bf C}_{\tilde{\epsilon}} + {\bf C}_{\tilde{n}} \right)^{-1} \hat{\bf G},
\label{Eq23a3}
\end{equation}
where ${\bf C}_{\tilde{\epsilon}} $ is one of the sub-blocks along the block diagonal of ${\bf C}_{\epsilon} $ in~(\ref{Eq21}). As in the case of ZF, the form of the LMMSE receiver does not lend itself to calculation of closed-form expressions for the expectations required to evaluate~(\ref{Eq25}), and hence in the next section we numerically evaluate the lower bound on the achievable rate for this approach.

\begin{figure*}[htb]
    \centering
  {{\includegraphics[width=7.2in]{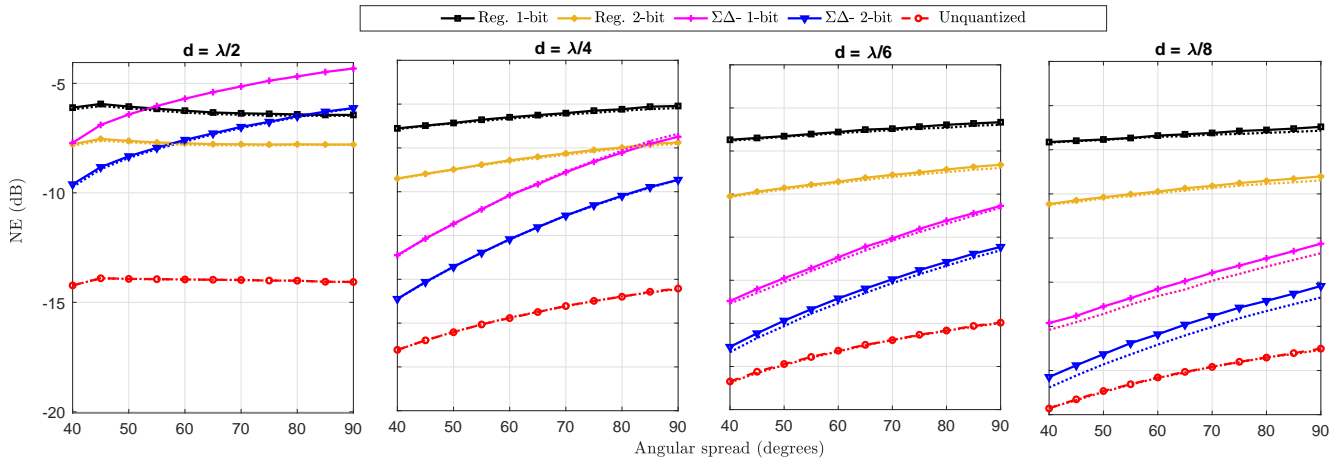} } \vspace*{-5pt}
    \caption{NE of channel estimates for different angular spreads of user AoAs and inter-element antenna spacings. Dotted lines indicate the NE of the channel estimate when there is no mutual coupling.}
\label{fig:3}}
    \end{figure*}  

\begin{figure}[h!]%
    \centering
  {{\includegraphics[width=3.7in]{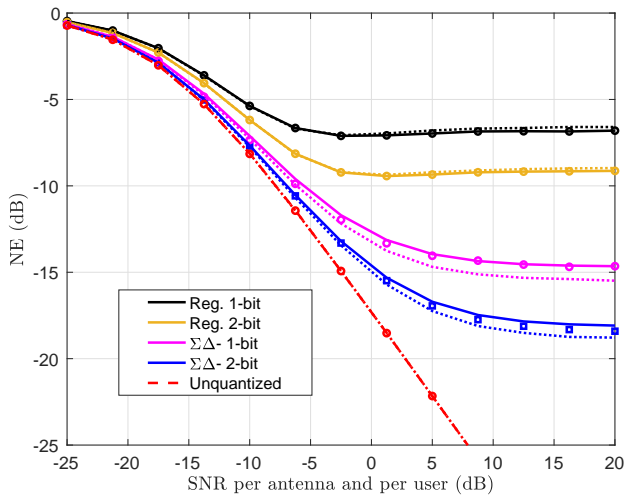} } 
    \caption{NE of channel estimate, $M=128, N=K=10$. The theoretical values are indicated by solid lines whereas the simulated values are denoted by circles. {\color{black} Dotted lines show the performance when mutual coupling is removed}.}
\label{fig:2}}
    \end{figure}   

\section{Simulation Results}
\label{SimResults}
In this section, we numerically evaluate the NE and sum spectral efficiency achieved with the $\Sigma \Delta$ massive MIMO system using one or two-bit outputs. The inter-element spacing is defined by $\delta$, and unless otherwise indicated, the simulations will employ a uniform linear array (ULA) equipped with $128$ antennas. The number of pilot symbols and number of users are both 10 ($N=K=10$), and orthogonal pilot sequences are used based on the $N \times N$ DFT matrix. We further assume that the downlink transmission power is equal to the pilot transmission power, i.e. $\rho_d = \rho$.

The users are assumed to be located within a sector centered on the broadside of the antenna array, so the steering angle of the $\Sigma\Delta$ array is set to $\psi=0\degree$, and $\mathcal{S}_{\theta}=[-\frac{\Theta}{2},\frac{\Theta}{2}]$. {\color{black} We assume the spatial covariance matrix is the same for all users, and defined by $L=50$ uniformly spaced signal arrivals $\theta_l$} in the interval defined by $\mathcal{S}_{\theta}$. The parameters used in modeling the mutual coupling of the array are: $R = 50~\ohm$, $T_A = 290~\mathrm{K}$, $\varrho_n = 0$, $B = 20~\mathrm{MHz}$, $R_N = R$, $\sigma_i^2 = 2k_B T_A B/R$, and $\sigma_v^2 = 2k_B T_A BR$. The simulations performed in the absence of mutual coupling assume that ${\bf Z} =R {\bf I}$ and ${\bf T} = 0.5 {\bf I}$~\cite{pirzadeh_spawc}.

The NE of the channel estimate is evaluated over $500$ independent realizations of the channel. For two-bit ADCs, we choose the optimum levels $\{\nu_1, \nu_2, \nu_3, \nu_4 \}$ as per~\cite{max1960quantizing}. We also use the sum spectral efficiency as a performance measure defined as
\begin{equation}
   {\rm SE} = \frac{T-N}{T} \sum_{i=1}^K R_k, 
    \label{Eq}
\end{equation}
where $T$ is length of the coherence interval during which the channel remains constant. We assume that $T = 200$ symbols.

We compare the performance of the $\Sigma\Delta$ LMMSE channel estimator derived above with the one-bit Bussgang LMMSE (BLMMSE) channel estimator of~\cite{Li_channel} and the LMMSE channel estimate for standard two-bit quantization. The LMMSE algorithm for the two-bit case can be derived by combining the analysis of~\cite{Li_channel} with that in Sections~\ref{TwoBit} and~\ref{ChEst} by replacing $r_m$ with $x_m$. The NE of an LMMSE channel estimator using unquantized measurements is also evaluated. Fig.~\ref{fig:3} shows the estimation performance as a function of the angular spread $\Theta$ of the user AoAs for four different antenna spacings. As expected, the performance of the spatial $\Sigma\Delta$ approach improves as either the antenna spacing $\delta$ or the size of the users' angular spread $\Theta$ decreases. Without oversampling, {i.e.}, when $\delta=\lambda/2$, the $\Sigma\Delta$ array offers no benefit over regular one- and two-bit quantization, except for a very narrow region near broadside. For $\delta \le \lambda/4$, however, there is a significant gain for angular sectors up to $90^\circ$ and beyond. {\color{black} The dotted lines in the plots show the performance of the $\Sigma\Delta$ channel estimator if the effect of mutual coupling is removed. While mutual coupling increasingly degrades the channel estimation performance as the antenna spacing decreases, the overall effect is not large.} For the remaining numerical examples, we will set $\Theta=60\degree$ and $\delta=\lambda/6$.

\begin{figure}[h!]%
    \centering
  {{\includegraphics[width=3.7in]{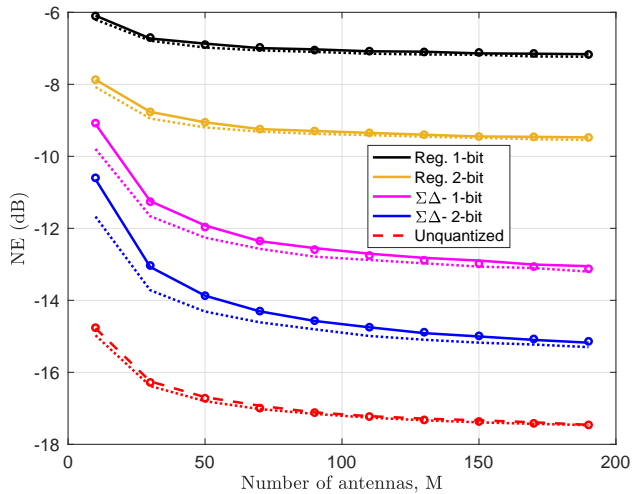} } 
    \caption{NE of channel estimate as a function of the number of antennas, ${\rm SNR} = 0$dB. Dotted lines indicate the NE of the channel estimate when there is no mutual coupling.}
\label{fig:4}}
    \end{figure}

Fig.~\ref{fig:2} shows the NE of the channel estimates as a function of SNR with $M=128$ antennas. The solid lines show the NE predicted by~(\ref{Eq20}), the symbols indicate the simulation {\color{black} results, and the dotted lines show the performance without the effect of mutual coupling} (this convention will be followed in all subsequent plots). We see that there is excellent agreement between our theoretical expression and the simulations. At low-to-medium SNRs, the performance of the $\Sigma\Delta$ channel estimates is very close to that of the unquantized MMSE channel estimate. The NE with two-bit $\Sigma\Delta$ ADCs is also lower than that achieved by one-bit $\Sigma\Delta$ ADCs beyond an SNR of $-5$dB. The gap between the two widens as the SNR 
increases and the error floor is about $-15$dB with one-bit $\Sigma\Delta$ ADCs and $-18$dB with two-bit $\Sigma\Delta$ ADCs. It is seen that LMMSE channel estimation with the $\Sigma\Delta$ array offers a significant advantage over the conventional one-bit and two-bit quantized arrays. The error floor of the $\Sigma\Delta$ channel estimators is around 8-9dB lower than their counterparts employing standard quantization. The advantage of the spatial $\Sigma\Delta$ approach is further seen in Fig.~\ref{fig:4} where performance is plotted as a function of $M$ for a fixed SNR of 0dB. {\color{black} These results indicate that the loss due to mutual coupling diminishes as the size of the array grows}.
   
\begin{figure}
\begin{subfigure}{.5\textwidth}
  \centering
  \includegraphics[width=3.7in]{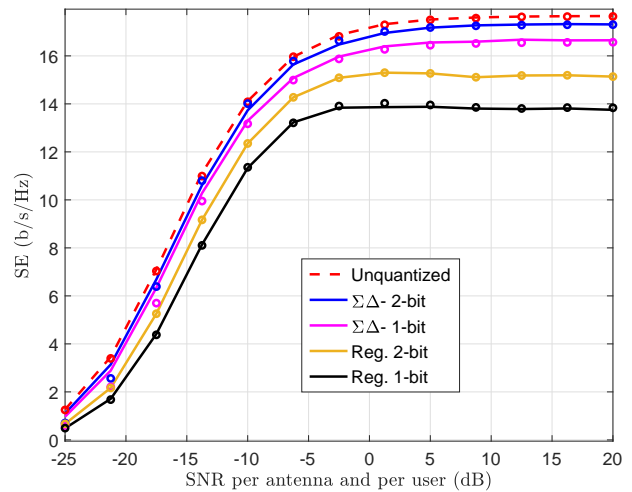}  
  \caption{}
\end{subfigure}
\begin{subfigure}{.5\textwidth}
  \centering
  \includegraphics[width=3.7in]{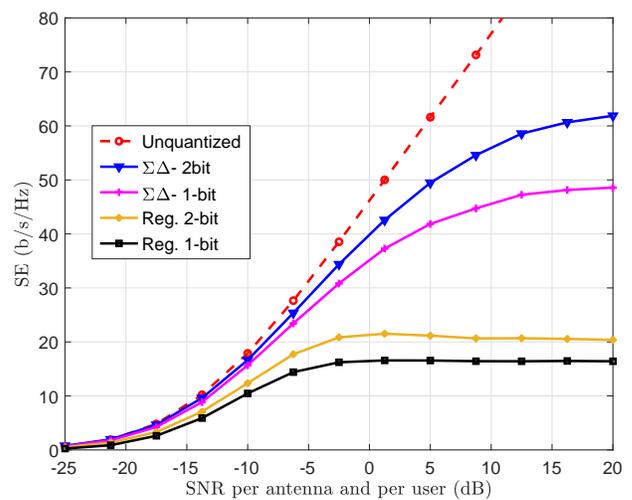}  
  \caption{}
\end{subfigure}
\begin{subfigure}{.5\textwidth}
  \centering
  \includegraphics[width=3.7in]{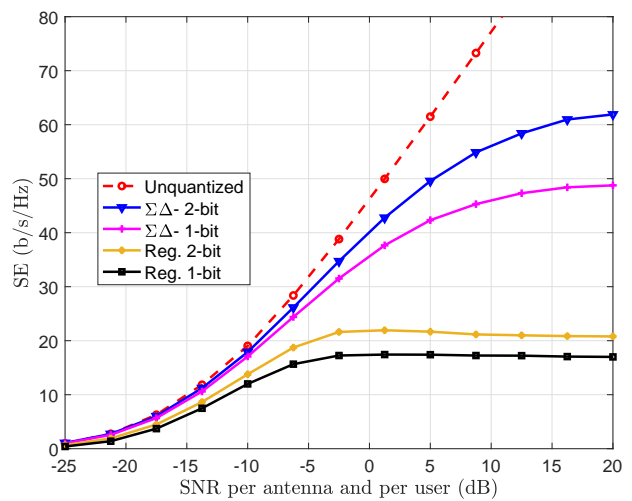}  
  \caption{}
\end{subfigure}
\caption{Sum spectral efficiency (SE) with (a) MRC (b) ZF (c) MMSE receivers ($M=128, K=10$). }
\label{fig5}
\end{figure}
 
In Fig.~\ref{fig5}, we plot the theoretical and simulated sum spectral efficiency (SE) achieved by the MRC and ZF receivers using the LMMSE channel estimate. Fig.~\ref{fig5}(a) shows that the SE of the $\Sigma\Delta$ architectures is close to that of an unquantized system and a bit higher than that of the one-bit massive MIMO system, although for MRC the difference in SE is not so large since multi-user interference is more of a limiting factor. There is excellent agreement between the simulations and our analytical expression in~(\ref{Eq26}). More impressive results are obtained for the case of a ZF receiver, as shown in Fig.~\ref{fig5}(b). At high SNRs, the throughput achieved with two-bit $\Sigma\Delta$ ADCs is around $60$ bits/s/Hz, almost 2.5 times that achieved with regular two-bit ADCs. With one-bit $\Sigma\Delta$ ADCs, the maximum throughput is around $50$ bits/s/Hz, also around $2.5$ times that achieved with regular one-bit ADCs. Of course, there is also a much bigger gap between the spectral efficiencies achievable by the $\Sigma\Delta$ and unquantized systems as well, especially at high SNR. From Fig.~\ref{fig5} (c), we see that the performance of the MMSE receiver is nearly identical to that of the ZF receiver since the number of users is fairly small ($K=10$).

\begin{figure}
  \includegraphics[width=3.6in]{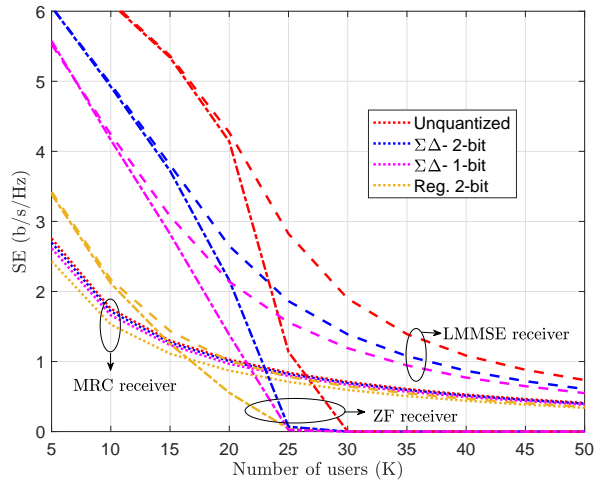}  
\caption{Per-user achievable rate as a function of the number of users, SNR = $5$dB, $M=128$. Dotted lines correspond to MRC, dash-dot lines for ZF, and dashed lines for LMMSE.}
\label{fig7}
\end{figure}

Finally, Fig.~\ref{fig7} shows a plot of the average per-user achievable rate for the MRC (dotted lines), ZF (dash-dotted lines) and LMMSE (dashed lines) receivers versus the total number of users. For MRC, the per-user rate achieved with both the one-bit and two-bit $\Sigma\Delta$ ADC arrays is essentially identical to that achieved without quantization. The numerator term of the logarithm in~(\ref{Eq26}) decreases with $K$ while the inter-user interference term in the denominator increases with $K$. The net effect is that, as the number of uplink users increases, the effective SNR per user for MRC decreases as $O(K^2)$. When $K$ is not too large, the per-user rate for ZF is significantly higher than for MRC, with the $\Sigma\Delta$ architecture falling in between the standard one-bit and unquantized systems. For an average per-user rate of 2 bits/s/Hz, MRC can support 7-8 users almost independently of the quantization level, while for ZF/MMSE the number of users increases to 10 for standard two-bit quantization, and 16-24 for the $\Sigma\Delta$ array. As the number of users increases, the channel matrix becomes ill-conditioned and the sum rate achieved with the ZF receiver tends towards zero. This is primarily due to the fact that the users' signals are confined to arrive from a $60^\circ$ angular sector. However, the LMMSE receiver is more robust and continues to provide a positive, albeit decreasing, average rate for the users even as $K$ is increasing.

\section{Conclusion}
In this paper, we considered channel estimation in massive MIMO systems employing spatial $\Sigma\Delta$ modulation with one- or two-bit ADCs. We used an element-wise Bussgang decomposition to derive a new LMMSE channel estimator that
takes into account the effect of the correlation between the quantizer outputs. We further analyzed the uplink rate that is achievable for MRC, ZF and LMMSE receivers implemented with the LMMSE channel estimate. The simulation results show that, in situations where the users are confined to a certain angular sector or the array elements are more closely spaced than one-half wavelength, the spatial $\Sigma\Delta$ approach is able to achieve
significantly better channel estimates and spectral efficiency than systems employing direct quantization using one- and two-bit ADCs. At low-to-medium SNR values, the performance gap between the $\Sigma\Delta$ array and a system with infinite-resolution ADCs is negligible. We derived a theoretical expression for the channel estimation error and showed that it provides excellent predictions for the performance obtained numerically. We also derived lower bounds on the achievable spectral efficiency. Overall, our results indicate that the spatial $\Sigma\Delta$ architecture provides a promising approach for increasing the energy efficiency and reducing the hardware complexity and fronthaul throughput of large-scale antenna systems.

\section*{Appendix A}

We will prove that $ \mathbb{E}\left[x_m q_n^* \right] \approx 0, \, \forall m,n$ following an approach similar to that in Appendix A of~\cite{Li_channel}. We can express $ \mathbb{E}\left[x_m q_n^* \right]$ as
\begin{equation}
 \mathbb{E}\left[x_m q_n^* \right] = \mathbb{E}_{r_n} \left[  \mathbb{E}\left[x_m q_n^* | r_n \right] \right] = \mathbb{E}_{r_n} \left[  \mathbb{E}\left[x_m | r_n \right] q_n^*  \right],
    \label{EqAppA1}
\end{equation}
where the last equality follows from the fact that $q_n = y_n - r_n$ is fixed for a given $r_n$. $\mathbb{E}\left[x_m | r_n \right]$ is the MMSE estimate of $x_m$ given $r_n$. Since we assume that $r_n$ is approximately a Gaussian random variable, the MMSE estimate of the Gaussian variable $x_m$ will be approximately equivalent to the LMMSE estimate given by
\begin{equation}
   \mathbb{E}\left[x_m | r_n \right] \approx \frac{\mathbb{E}\left[ x_m r_n^* \right]}{\sigma_{r_n}^2} r_n.
    \label{EqAppA2}
\end{equation}
Using~(\ref{EqAppA2}) and the fact that $r_n$ is uncorrelated with $q_n$, we have
\begin{equation}
   \mathbb{E}\left[x_m q_n^* \right]  \approx \frac{\mathbb{E}\left[ x_m r_n^* \right]}{\sigma_{r_n}^2} \mathbb{E}_{r_n} \left[   r_n q_n^*  \right] =0.
    \label{EqAppA3}
\end{equation}
Therefore, $\mathbb{E}\left[x_m q_n^* \right] \approx 0, \, \forall m,n$ and hence $\mathbb{E}\left[ {\bf x} {\bf q}^H \right] \approx {\bf 0}$.
\bibliographystyle{IEEEbib}
\bibliography{IEEEabrv,bibJournalList,refs}

\end{document}